# Layer thickness-dependent phonon properties and thermal conductivity of $MoS_2$

Xiaokun Gu[1], Baowen Li[1] and Ronggui Yang[1,2,*]

[1]Department of Mechanical Engineering, University of Colorado at Boulder, Colorado, 80309, USA

[2]Materials Science and Engineering Program, University of Colorado at Boulder, Colorado, 80309, USA

*ronggui.yang@colorado.edu

**Abstract**

For conventional materials, the thermal conductivity of thin film is usually suppressed when the thickness decreases due to phonon-boundary scattering. However, this is not necessarily true for the van der Waals solids if the thickness is reduced to only a few layers. In this letter, the layer thickness-dependent phonon properties and thermal conductivity in the few-layer $MoS_2$ are studied using the first-principles-based Peierls-Boltzmann transport equation approach. The basal-plane thermal conductivity is found to monotonically reduce from 138 W/mK to 98 W/mK for naturally occurring $MoS_2$, and from 155 W/mK to 115 W/mK for isotopically pure $MoS_2$, when its thickness increases from one layer to three layers. The thermal conductivity of tri-layer $MoS_2$ approaches to that of bulk $MoS_2$. Both the change of phonon dispersion and the thickness-induced anharmonicity are important for explaining such a thermal conductivity reduction. The increased anharmonicity in bi-layer $MoS_2$ results in stronger phonon scattering for $ZA_i$ modes, which is linked to the breakdown of the symmetry in single-layer $MoS_2$.

**Main text**

Single-layer and few-layer molybdenum disulfide ($MoS_2$) have offered a great opportunity to realize novel nanoelectronic and photonic devices, due to its unique physical properties.[1-3] Its layer-dependent bandgap, changing from a 1.2 eV indirect bandgap in a bulk sample to a 1.8 eV direct bandgap in a single-layer one,[4] renders the potential applications in switchable transistors[5] and sensitive photodetectors.[6, 7] In addition, the Seebeck coefficient and the electrical conductivity of $MoS_2$ was also reported to be tunable by its layer thickness. The maximum power factor in bi-layer $MoS_2$[8] was found to be comparable to state-of-the-art commercial thermoelectric material $Bi_2Te_3$. Apart from the electronic and optical properties, it is of practical importance to study phonon properties and thermal transport in few-layer $MoS_2$ due to the reliability and thermal management concerns of $MoS_2$-based electronics and photonics as well as its potential applications in thermoelectrics.

In the past two decades, considerable attention has been paid to the size effects of phonon transport in semiconductor nanostructures, such as thin films,[9] nanowires,[10-12] superlattices[13, 14] and nanocomposites.[15-18] In these semiconductor nanostructures, interfaces play a crucial role on reducing the thermal conductivity compared with their bulk counterparts by inducing additional interface scattering of phonons. However, such interface scattering might not be as important for the van der Waals solids, where two-dimensional layers are stacked together through very weak van der Waals bonding and the surface of few-layer two-dimensional materials can be atomically smooth when the thickness is reduced to only a few layers.[19] For example, both experiments[20] and numerical simulations[21-23] showed that the thermal conductivity of graphene gradually decreases when the layer number is increased and approach to the bulk value when there are four to five layers. It is generally believed that the thermal conductivity reduction in few-layer graphene is due to the breakdown of the selection rule that arises out of the reflection symmetry of the single-layer graphene.[24] However, some other two-dimensional materials with different crystal structures, such as $Bi_2Te_3$ and $TaSe_2$, exhibit quite different layer-dependence of the thermal conductivity from graphene. Qiu and Ruan[25] found a non-monotonic dependence of the thermal conductivity on layer thickness of $Bi_2Te_3$ from their equilibrium molecular dynamics simulations. The single-layer $Bi_2Te_3$ has the highest thermal conductivity, which is then reduced to the minimum value for the three-layer $Bi_2Te_3$, and then converged back to the bulk value. Yan et al.[26] measured the thermal conductivity of two-dimensional transition metal dichalcogenide $TaSe_2$ with 1T structure using the optothermal Raman measurement, and found that the thermal conductivity of these $TaSe_2$ films at room temperature decreases from its bulk value of 16 W/mK to 9 W/mK in 45-nm-thick films. These studies indicate that the crystal structures of the two-dimensional materials might play an important role in determining their layer thickness-dependent thermal conductivity. Because the crystal structure of $MoS_2$ is quite different from the materials studied before, it is unclear how the thermal conductivity of two-dimensional $MoS_2$ changes with its layer thickness.

Along this line of curiosity, many experimental works have been conducted to measure the thermal conductivity of single-layer, few-layer and bulk $MoS_2$. The measured results are 34.5 W/mK[27] and 84 W/mK[28] for single-layer one, 77 W/mK,[28] 46 W/mK,[29] 50 W/mK[29] and 52 W/mK[30] for 2-layer, 4-layer, 7-layer and 11-layer $MoS_2$, respectively. Liu *et al.*[31] reported a large thermal conductivity value for bulk $MoS_2$, around 100 W/mK. The details of these measurements are summarized in Table I. Direct comparison among these experimental data is challenging since both sample quality and experiment conditions are different from different research groups.

Theoretical work has also been employed to study the thermal conductivity in single-layer and few-layer MoS$_2$. A recent classical molecular dynamics simulations[32] using empirical interatomic potential showed that the thermal conductivity of MoS$_2$ is about 5 W/mK for all MoS$_2$ with different layer numbers. This calculated thermal conductivity is one to two orders-of-magnitude smaller than the measured results, which indicates that the anharmonicity of the intralayer interaction from the empirical potential is severely overestimated. On the other hand, the first-principles-based Peierls-Boltzmann transport equation (PBTE) method[33-38] has been employed to predict the thermal conductivity of many bulk and two-dimensional materials, including single-layer MoS$_2$,[39, 40] which showed satisfactory agreement with the experiment results.

In this letter, we apply the first-principles-based PBTE approach to study the dependence of thermal conductivity of two-dimensional MoS$_2$ on its layer thickness. Due to the computational power limitations, we calculate the thermal conductivity of one- to three-layer MoS$_2$, but compare them with bulk MoS$_2$. The thermal conductivity of single-layer MoS$_2$ is found to be the highest among all samples studied, and the thermal conductivity decreases with the thickness changing from one to three layers. Detailed phonon scattering rate analysis shows that the anharmonicity is significantly increased in the bi- and tri- layer MoS$_2$ compared with single-layer MoS$_2$, which suppresses the heat conduction ability of flexural acoustic phonons.

Figure 1(a) illustrates the unit cells for the few-layer MoS$_2$ as well as bulk MoS$_2$ studied in this work. The MoS$_2$ monolayers are placed parallel to the $x$-$y$ plane. To calculate the basal-plane thermal conductivity in few-layer and bulk MoS$_2$, we apply a small temperature gradient $\frac{dT}{dx}$ in the region between the two ends of the MoS$_2$ sheets with a distance $L$ apart in the $x$ direction, as shown in Fig. 1(b). The phonon population is perturbed and can be written as $n_{qs}\left(=n_{qs}^0+\right.$

$n_{qs}^0(n_{qs}^0 + 1)\frac{dT}{dx}F_{qs}^x$), where $n_{qs}^0$ denotes the equilibrium Bose-Einstein distribution, the subscript $qs$ stands for the $s$-th phonon mode with $\hbar q$ momentum ($\hbar$ is the Planck constant) in the first Brillouin zone, and $F_{qs}^x$ is deviation function. The deviation function can be solved from the linearized PBTE when the steady state of phonon flow is achieved. In this work, we consider the three-phonon scattering, the boundary scattering, and isotope scattering mechanisms. The details of PBTE, including its formalism, the expression of the scattering rates of different scattering mechanisms considered in the PBTE and the solution of PBTE, can be found in Section I of the Supplemental Information. Similarly, for the calculation of the cross-plane thermal conductivity of bulk $MoS_2$, the temperature gradient is imposed along the $z$ direction between two separated parallel regions, as illustrated in Fig. 1(c). We solve the corresponding PBTE and obtain the deviation function $F_{qs}^z$.

With the deviation function of each phonon mode obtained from the PBTE, the thermal conductivity can be expressed as the summation of the contributions from all phonon modes through

$$\kappa^{\alpha\alpha} = \frac{1}{(2\pi)^3}\sum_s \int \hbar\omega_{qs} v_{qs}^\alpha n_{qs}^0(n_{qs}^0 + 1)F_{qs}d\bm{q}, \qquad (1)$$

where $\alpha$ stands for $x$ or $z$ (basal-plane or cross-plane), $\omega_{qs}$ and $v_{qs}$ are the frequency and group velocity of phonon mode $qs$, which can be straightforwardly obtained from phonon dispersion relation.

To obtain the information of phonon frequency, group velocity, equilibrium phonon distribution function as well as the phonon-phonon scattering rates for computing the thermal conductivity, both the second-order harmonic force constants and the third-order anharmonic force constants of bulk $MoS_2$ and few-layer $MoS_2$ are required. We perform a series of first-principles calculations based on the local density approximation (LDA),[41] which has been widely employed to study the thermal conductivity of many van der Waals layered crystals, including graphite,[42] $Bi_2Te_3$[43] and SnSe,[44] to extract these force constants. The information of first-principles calculations to determine the crystal structures and to extract the interatomic force constants are also summarized in Section II of the Supplemental Information. The optimized in-plane and cross-plane lattice constants of bulk $MoS_2$, $a_0$ and $c_0$, are 3.14 Å and 12.05 Å, which are close to the experimental values of 3.15 Å and 12.3 Å.[45] Single-layer and few-layer $MoS_2$ have the same in-plane lattice constants as bulk $MoS_2$.

Since phonon properties and the calculated thermal conductivity are very sensitive to the interatomic force constants, we perform the following tests to validate our extracted force constants. Figure 2 shows the calculated phonon dispersion curves of bulk $MoS_2$ using the extracted harmonic force constants, which are in reasonable agreement with the available experimental data of from the inelastic neutron scattering measurement.[45]

We then turn to the verification of the third-order anharmonic force constants. Recent studies have shown that the cutoff of anharmonic interaction is crucial to the calculation of thermal conductivity.[46] While too small a cutoff tends to overestimate the thermal conductivity, a larger cutoff requires much more computational resources. To make our calculations affordable, we carefully chose the range of anharmonic interaction to be 6 Å. We validate the extracted third-order force constants by calculating the mode-specific Gruneisen parameters, which serve as indicators for the degree of anharmonicity of the crystal, using bulk $MoS_2$ as a testing case. The mode-specific Gruneisen parameters can be calculated in two ways. The finite difference approach gives,

$$\gamma_{\mathbf{q}s} = -(d\omega_{\mathbf{q}s}/dV)/(\omega_{\mathbf{q}s}/V), \tag{3}$$

with the crystal volume $V$.

From the perturbation theory using the third-order force constants as inputs, the mode-specific Gruneisen parameter is expressed as[35, 47]

$$\gamma_{\mathbf{q}s} = \frac{1}{6\omega_{\mathbf{q}s}^2}\sum_{\tau,\mathbf{R}'\tau',\mathbf{R}''\tau''}\sum_{\alpha\beta\gamma}\Psi^{\alpha\beta\gamma}_{0\tau,\mathbf{R}'\tau',\mathbf{R}''\tau''}\frac{\varepsilon^{\tau\alpha}_{\mathbf{q}s}\varepsilon^{\tau'\beta}_{\mathbf{q}'s'}}{\sqrt{M_\tau M_{\tau'}}}\exp(i\mathbf{q}\cdot\mathbf{R}')\,r^{\gamma}_{\mathbf{R}''\tau''}, \qquad (4)$$

where $(\mathbf{R}, \tau, \alpha)$ refers to the degree of freedom corresponding to the α direction of the τ-th basis atom in the unit cell located at position $\mathbf{R}$, $\Psi$ is the third-order anharmonic force constant, ε and $\mathbf{r}$ are the polarization vector component and equilibrium atomic position.

Figure 3 shows the calculated mode-specific Gruneisen parameters of acoustic and low-lying optical (below the frequency gap at around 250 cm$^{-1}$) phonon modes for bulk MoS$_2$. These phonon modes accounts for more than 90% of its total thermal conductivity. Clearly, the results from these two computing methods are consistent with each other, confirming that our choice on the cutoff for the anharmonic interaction indeed gives converged results.

With the phonon dispersion calculated from the second-order harmonic force constants and the phonon-phonon scattering rates calculated using the extracted third-order anharmonic force constants, we calculate the thermal conductivity by solving the PBTE. We use 25×25×7 and 65×65×1 q-points to sample the reciprocal space for phonon scattering and thermal conductivity calculations respectively, which ensure the presented basal-plane thermal conductivity data be convergent to 2% if the meshes are further refined. To report the thermal conductivity of $N$-layer MoS$_2$, we assign it with a thickness of $N/2$ times of the cross-plane lattice constants of bulk MoS$_2$ calculated from first-principles calculations, i.e., 6.025 Å, 12.05 Å and 18.075 Å for single-layer, bi-layer and tri-layer MoS$_2$, respectively.

Figure 4 shows the calculated basal-plane thermal conductivity of an infinitely large bulk MoS$_2$ as a function of temperature, in comparison with the available experiment data. The basal-plane thermal conductivity of bulk MoS$_2$ made up of the naturally occurring Mo and S isotopes decreases from 340 W/mK to 73 W/mK when the temperature increases from 100 K to 400 K, while the cross-plane thermal conductivity changes from 11.4 W/mK to 2.6 W/mK. At room temperature, the basal-plane thermal conductivity from our calculation is 98 W/mK, which falls in the range of 85-112 W/mK obtained from a recent pump-probe measurement.[31] The thermal conductivity of isotopically pure MoS$_2$ was also shown in Fig. 4. Due to the absence of isotope scattering, the basal-plane thermal conductivity is 117 W/mK in the isotopically pure crystal, about 20% higher than the naturally occurring one.

Figure 5 shows the cross-plane thermal conductivity of bulk MoS$_2$. The cross-plane thermal conductivity of the naturally occurring MoS$_2$ is calculated to be 3.5 W/mK at room temperature. We notice that the measured cross-plane thermal conductivity ranges from 2 W/mK to 2.5 W/mK using ultrafast laser-based pump-and-probe measurements, which is consistently lower than the calculated value.[31, 48] This indicates that sample quality and measurement geometry could play an important role in the determination of the measured cross-plane thermal conductivity.

One possible explanation is that the thermal conductivity extracted from the pump-and-probe measurements is dependent on the modulation frequency.[49] The pump beam with a modulation frequency of $\omega_{\text{pump}}$ could only heat a limited region beneath the heating surface where the depth of this heating region can be roughly estimated as the penetration depth $l_p = \sqrt{\kappa/\pi C \omega_{\text{pump}}}$ with the heat capacity of the material $C$ (=1.89 J/cm$^3$K for bulk MoS$_2$). When the mean free path of a phonon is smaller than the penetration depth $l_p$, it is likely to transport ballistically across the heating region, leading to a suppression of its contribution to the total thermal conductivity. With the common modulation frequency $\omega_{\text{pump}}$ ($\approx$ 10 MHz) [31] employed in these experiments, along with the estimated cross-plane thermal conductivity of bulk $MoS_2$, 2-5 W/mK, the penetration depth is only about 200-300 nm. This indicates that the heat carried by the phonons with mean free paths larger than ~250 nm is suppressed in the pump-probe measurements. As a result, the deduced thermal conductivity from the pump-and-probe measurement has a lower value than its true thermal conductivity. To evaluate the contribution from these long mean free path phonons, we calculate the cross-plane thermal conductivity of $MoS_2$ by leaving the two reservoirs 250 nm apart, mimicking the phonon transport across the heating region in the pump-and-probe measurements. We find that the thermal conductivity from this simulation with short distance is about 2.1 W/mK, which indeed fall into the range of the measured cross-plane thermal conductivity of $MoS_2$.

After the validation of the theoretical calculations on bulk $MoS_2$ with the insights to the pump-and-probe measurements, we study the thermal conductivity of single and few layer $MoS_2$. To mimic the measurement conditions for the in-plane thermal conductivity, where the characteristic size of the suspended samples is usually in the order of several microns, we impose a sample size of 10 μm to include boundary scattering to solve the PBTE. Figure 6 shows the calculated thermal conductivity of $MoS_2$ as a function of layer number at room temperature, in comparison with the recent measurement results.[27-31] For naturally occurring $MoS_2$, the calculated thermal conductivity values are 138 W/mK, 108 W/mK, 98 W/mK and 94 W/mK for single-layer, bi-layer, tri-layer and bulk samples, respectively. For the isotopically pure samples, the thermal conductivity values are consistently higher at 155 W/mK, 125 W/mK, 115 W/mK, and 112 W/mK, respectively. It is evident that the thermal conductivity of few-layer $MoS_2$, both naturally occurring and isotopically pure, decrease with the thickness from single layer to three layer, and the thermal conductivity of the tri-layer $MoS_2$ almost approaches to the bulk value.

To understand the mechanisms of thermal conductivity reduction from single-layer $MoS_2$ to thicker ones, we examine the contributions from the changed phonon dispersion curves of $MoS_2$ with one to three layers. Figure 7 shows the dispersion curves of acoustic and low-frequency optical branches ($< 250$ cm$^{-1}$), which together contribute to more than 90% of the total thermal conductivity. In the single-layer $MoS_2$, there are three acoustic branches, including one longitudinal acoustic (LA) branch, one transverse acoustic (TA) branch and one flexural acoustic (ZA) branch, whose frequencies becomes zero as the wavenumber $q$ approaches to zero, shown as the black solid curves in both Fig. 7(a) and 7(b). For $N$-layer $MoS_2$ (N=2,3), there are $3N$-3 low-frequency optical phonon branches beneath the frequency gap around 250 cm$^{-1}$ in addition to the 3 acoustic branches due to more basis atoms involved. It is clearly seen that the three groups of phonon branches, each of which involves $N$ branches are almost degenerated/overlapped with each other for the region away from the first Brillouin zone center, which is similar to the observation on multi-layer graphene. We denote these phonon branches in multi-layer $MoS_2$ as LA$_i$, TA$_i$ and ZA$_i$ branches ($i = 1,.., N$), where LA$_i$ (TA$_i$ and ZA$_i$) are sorted ascendingly according to the phonon frequency at $q = 0$.

For multi-layer graphene, the $N$ branches in each groups become nondegenerated near the first Brillouin zone center.[21, 22, 50] While the acoustic branch has a zero frequency at the zone center, the other $N$-1 (optical) branches gradually become less dispersive (for TA$_i$ and LA$_i$ branches) or even flat (for ZA$_i$ branches) as $q \to 0$. Unlike multi-layer graphene, the phonon dispersion in multi-layer $MoS_2$ are dramatically different. Two acoustic branches and $N$-2 optical branch are found in the group of ZA$_i$ while one acoustic branch and $N$-1 optical branches in the group of TA$_i$, and even more interestingly, all $N$ LA$_i$ are optical modes. In Fig. 7(a) and (b), we also observe kinks, indicated by arrows, occur at $q \approx 0.05$ ($2\pi/a_0$) on the ZA$_i$ branches, changing the shapes of $ZA_i$ and $LA_i$ branches. Near each kink, the corresponding $ZA_i$ branch experiences a flat-to-dispersive transition as $q$ decreases. The $ZA_2$ even becomes an acoustic branch with a large sound velocity due to the transition. In contrast, $LA_i$ branches turn from dispersive to flat. As a result, the $LA_2$ and the $ZA_2$ in bi-layer $MoS_2$, as well as the $LA_1$ and the $ZA_2$, or the $LA_3$ and the $ZA_3$ in tri-layer $MoS_2$, do not cross each other. Similar phenomena have been also reported in other materials,[51, 52] and are called avoided-crossing in literature.[51] For example, in some caged structures encapsulating guest atoms, such as clathrates,[51] a dispersive acoustic-phonon branch and a flat branch corresponding to the movement of guest atom do not cross each other but transit to be flat and dispersive, respectively. The avoided-crossing in the few-layer $MoS_2$ reduces the group velocity of the acoustic phonon modes nearby. In addition to reduced group velocities of the phonon modes near the avoided-crossing, the long-wavelength optical phonon modes in few-layer $MoS_2$ generally have slightly smaller group velocities than the acoustic phonons, reducing the average group velocity for the heat carrying phonons compared with the single-layer $MoS_2$.

One might expect that such changes in the phonon dispersion from one-layer to multi-layer $MoS_2$ could lead to significant change on thermal conductivity, due to the lower group velocity. To test this hypothesis, we re-calculate the thermal conductivity of bi-layer and tri-layer $MoS_2$ using their own phonon dispersion as shown in Figure 7, but ignoring the third-order anharmonic force constants corresponding to the interlayer interaction and assign their intralayer anharmonic force constants with the anharmonic force constants of single-layer $MoS_2$. The recalculated thermal conductivity of bi-layer and tri-layer $MoS_2$ is only 12% and 15% lower than the single-layer one, respectively, which is definitely smaller than 22% and 29% obtained from the calculations using their own anharmonic force constants. Apparently, the change in phonon dispersion is an important factor to reduce the thermal conductivity from a higher value at the single layer to the multi-layer MoS$_2$. However, phonon dispersion change is not the single factor. The change of anharmonic force constants could be as important in reducing the thermal conductivity values from single-layer to multi-layer MoS$_2$.

Figure 8 shows the phonon scattering rate, or the inverse of phonon lifetime, of the acoustic and low-lying optical modes for single-layer and bi-layer MoS$_2$ along the Γ-K direction, calculated using their respective third-order anharmonic force constants. Clearly the scattering rates for the in-plane phonon modes (TA$_i$ and LA$_i$) in bi-layer MoS$_2$, except the modes near the zone center, are almost unchanged compared with their counterparts (TA$_1$ and LA$_1$) in single-layer MoS$_2$. Since there is only a small fraction of phonons near the zone center, the total thermal conductivity of TA$_i$ and LA$_i$ branches should be very close to that of TA$_1$ and LA$_1$ in single-layer MoS$_2$.

In comparison, the scattering rates of ZA$_i$ phonons in bi-layer MoS$_2$ are significantly larger than ZA$_1$ phonons in single-layer MoS$_2$ throughout the entire first Brillouin zone. This could be understood by closely examining the third-order anharmonic force constants. Assuming that $n$, $m$ and $l$ refer to any Mo atoms in the same layer, the force constants, $\psi_{n,m,l}^{z,z,z}$, which are the third-order derivatives of the total energy of the crystal with respect to the $z$ coordinates of atom $n$, $m$ and $l$, are found to be zero in single-layer MoS$_2$ because of the mirror symmetry. This means that there is no anharmonicity induced by the relative motion among the three Mo atoms along z direction. Since the dominating atomic motion of ZA phonons is along z direction, the scattering rates of these ZA phonon modes are thus very small. When two MoS$_2$ monolayers are in contact with each other, the mirror symmetry breaks down. As a result, the third-order anharmonic force constants $\psi_{n,m,l}^{z,z,z}$ become non-zero, leading to stronger scattering of ZA$_i$ phonons in bi-layer

MoS$_2$. In addition to the third-order anharmonic force constants involving the out-of-plane motion of three Mo atoms, the third-order anharmonic force constants corresponding to the interlayer interaction also contributes to the anharmonicity, which is absent in the single-layer MoS$_2$.

The observation of low scattering rate of ZA modes in single layer MoS$_2$ induced by mirror symmetry is indeed similar to graphene. In graphene, the scattering with odd number of out-of-plane modes, such as ZA + TA -> LA and ZA + ZA -> ZA, are totally prohibited due to its mirror symmetry.[24] However, there is still a notable difference between MoS$_2$ and graphene. Because the atom vibration of ZA modes also involves S atoms, the forbidden scattering channels in graphene are not totally forbidden in MoS$_2$. Therefore, while the thermal conductivity reduction from single-layer graphene to bi-layer graphene is mainly attributed to the stronger phonon scattering in bi-layer graphene due to the breakdown of the symmetry selection rule, which accounts for 70% reduction, both the change of phonon dispersion and the enhanced phonon scattering strength are important for explaining the thermal conductivity reduction from single layer to multilayer MoS$_2$.

In summary, we study the layer thickness-dependent phonon properties and thermal conductivity in the few-layer MoS$_2$ using first-principles-based PBTE calculations. The basal-plane thermal conductivity is found to monotonically reduce from 138 W/mK to 98 W/mK for naturally occurring MoS$_2$, and from 155 W/mK to 115 W/mK for isotopically pure MoS$_2$, when its thickness increases from one layer to three layers. The thermal conductivity of tri-layer MoS$_2$ approaches to that of bulk MoS$_2$. Both the change of phonon dispersion and the thickness-induced anharmonicity are important for explaining such a thermal conductivity reduction. The increased anharmonicity in bi-layer $MoS_2$ results in stronger phonon scattering for $ZA_i$ modes, which is linked to the breakdown of the symmetry in single-layer $MoS_2$.

**Acknowledgements**

This work was supported by the National Science Foundation (Award No. 1511195) and DOD DARPA (Grant No. FA8650-15-1-7524). XG acknowledges the Teets Family Endowed Doctoral Fellowship. This work utilized the Janus supercomputer, which is supported by the National Science Foundation (Award No. CNS-0821794), the University of Colorado at Boulder, the University of Colorado at Denver, and the National Center for Atmospheric Research. The Janus supercomputer is operated by the University of Colorado at Boulder.

**Table:**

**Table I.** Experimental studies on the thermal conductivity in MoS$_2$. χ is the temperature coefficient of Raman signal and α is the absorption ratio used for data fitting. The details of measurements are summarized in a recent review (Ref. [19]).

| Ref. | Method | Sample type | Room-temperature thermal conductivity (W/mK) | Experimental conditions |
|---|---|---|---|---|
| Yan [27] | Raman | Exfoliated, transferred | 34.5 ±4 (1-Layer) | A$_{1g}$ mode, χ =0.011 cm$^{-1}$/K, α=9±1% ,170 nm diameter laser spot, suspended on 1.2-μm-diameter holes, ambient condition |
| Zhang [28] | Raman | Exfoliated, transferred | 84 ±17 (1-Layer) | A$_{1g}$ mode, χ =0.0203 cm$^{-1}$/K, α=5.2±0.1%,460-620 nm diameter laser spot, suspended on 2.5-to-5.0-μm-diameter holes, ambient condition |
| Zhang [28] | Raman | Exfoliated, transferred | 77 ±25 (2-Layer) | A$_{1g}$ mode, χ =0.0136 cm$^{-1}$/K, α=11.5±0.1%,460-620 nm diameter laser spot, suspended on 2.5-to-5.0-μm-diameter holes, ambient condition |
| Jo [29] | Micro-bridge | Exfoliated, transferred | 44–50 (4-Layer) | Suspended sample; length: 3 μm, width: 5.2 μm. |
| Jo [29] | Micro-bridge | Exfoliated, transferred | 48-52 (7-Layer) | Suspended sample; length: 8 μm, width: 2.2 μm. |
| Sahoo [30] | Raman | CVD, transferred | 52 (11-Layer) | A$_{1g}$ mode, χ= 1.23 × 10$^{-2}$cm/K, α = 10%,1-1.5 μm laser spot, suspended on a 10- μm-radius quadrant, ambient condition |
| Liu [31] | Pump-probe | Bulk | 85-112 | Modulation frequency of pump beam: 10.7 MHz |

**Figure captions:**

**Figure 1.** (a) Unit cells of single-layer, bi-layer, tri-layer and bulk $MoS_2$ for structure relaxation and phonon dispersion calculation. (b) Schematic of the simulation system used to calculate the basal-plane thermal conductivity of few-layer and bulk $MoS_2$. (c) Schematic of the simulation system to calculate the cross-plane thermal conductivity of bulk $MoS_2$.

**Figure 2.** Phonon dispersion of bulk $MoS_2$. Black lines are the results calculated using the second-order harmonic force constants from first-principles calculations. Blue dots are the experimental data from inelastic neutron scattering measurement.[45]

**Figure 3.** Mode-specific Gruneisen parameters of acoustic and low-lying optical phonons of bulk $MoS_2$. Blue dots are the mode-specific Gruneisen parameters calculated based on the definition using the finite-difference Red stars are the mode-specific Gruneisen parameters calculated based on the perturbation theory using the third-order anharmonic force constants.

**Figure 4.** Basal-plane thermal conductivity of bulk $MoS_2$ as a function of temperature. Blue dot is the experimental data from pump-probe measurement.[31]

**Figure 5.** Cross-plane thermal conductivity of bulk $MoS_2$ as a function of temperature. Blue and Green dots are the experimental data from pump-probe measurement.[31, 48]

**Figure 6.** Room-temperature basal-plane thermal conductivity of $MoS_2$ as a function of number of $MoS_2$ monolayers.

**Figure 7.** Phonon dispersion of few-layer MoS$_2$. Black lines are the dispersion for single-layer MoS$_2$. Blue, red and green lines refer to the ZA$_i$, TA$_i$ and LA$_i$ branches, respectively. Dash, dash-dot and short-dash lines refer to $i$=1, 2, 3, respectively.

**Figure 8.** Scattering rates of the acoustic and low-lying optical branches of phonons in single-layer and bi-layer MoS$_2$. Black lines are the scattering rates for single-layer MoS$_2$. Blue, red and green lines refer to the ZA$_i$, TA$_i$ and LA$_i$ branches, respectively. Dash and dash-dot lines refer to $i$ =1, 2, respectively.

**Figure 1.**

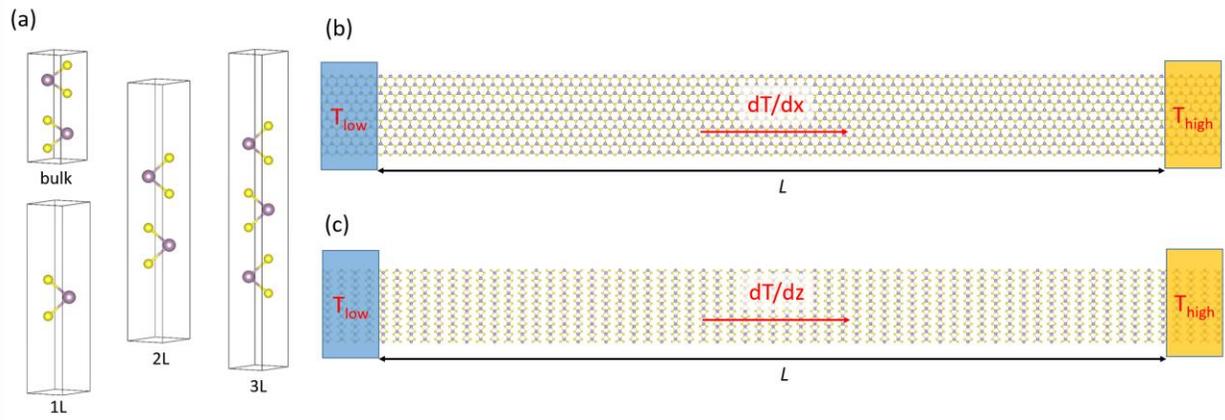

**Figure 2.**

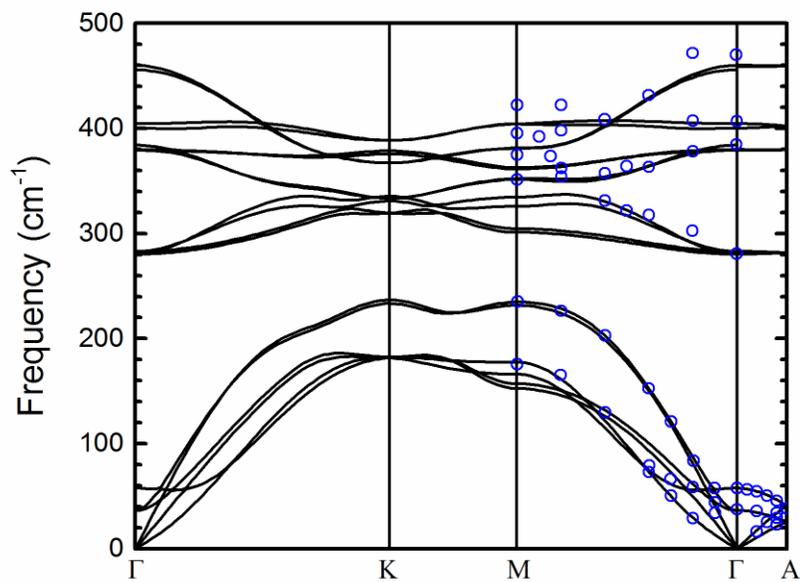

**Figure 3.**

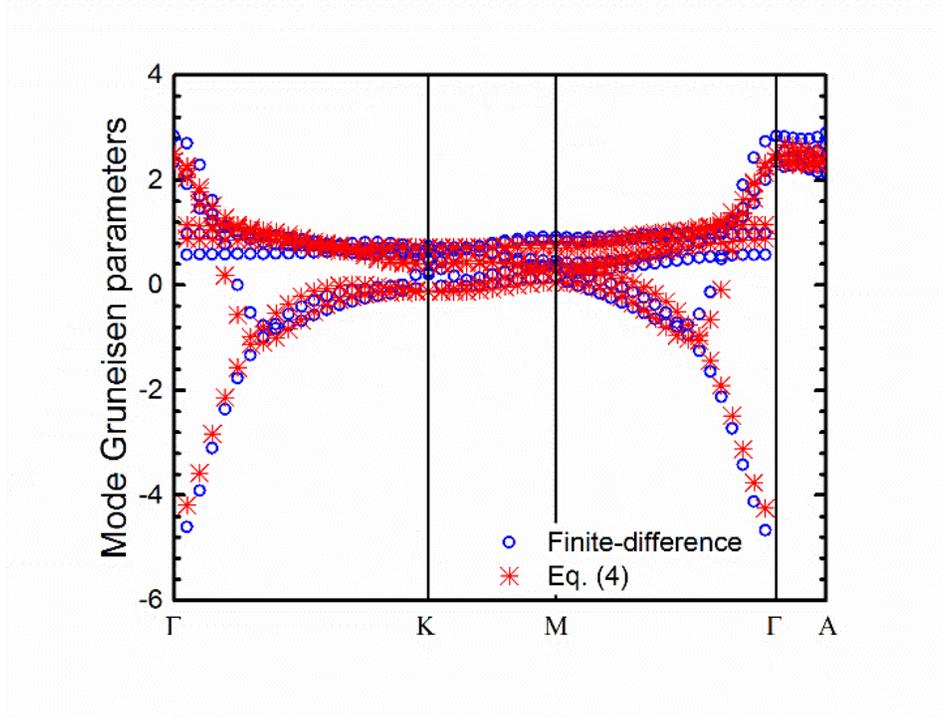

**Figure 4.**

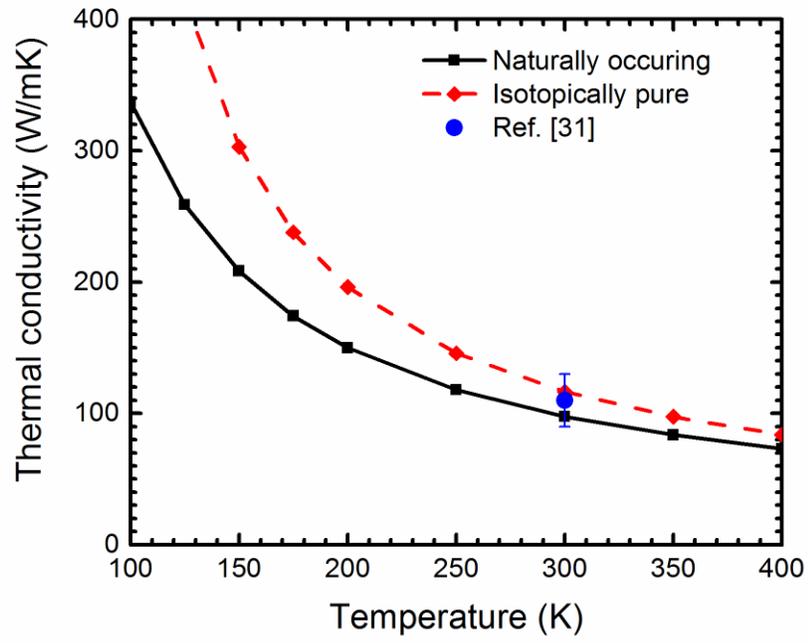

**Figure 5.**

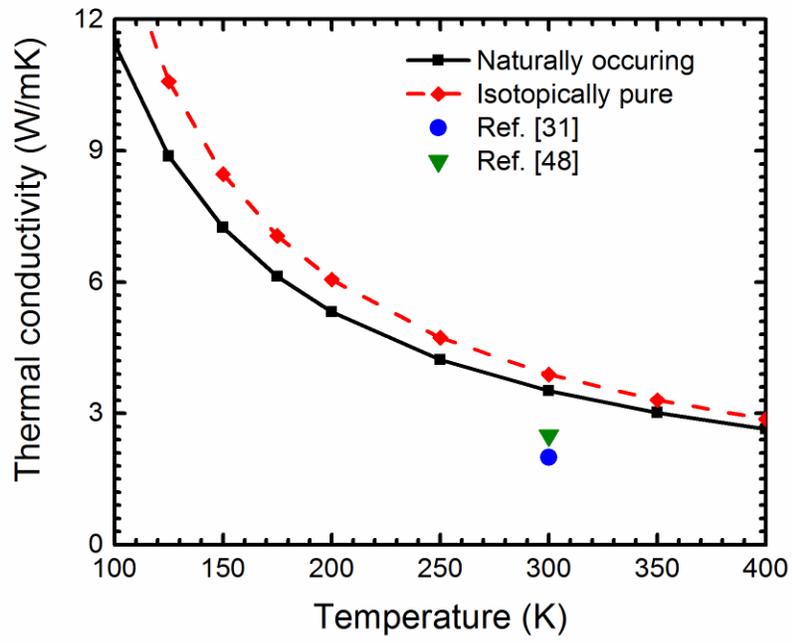

**Figure 6.**

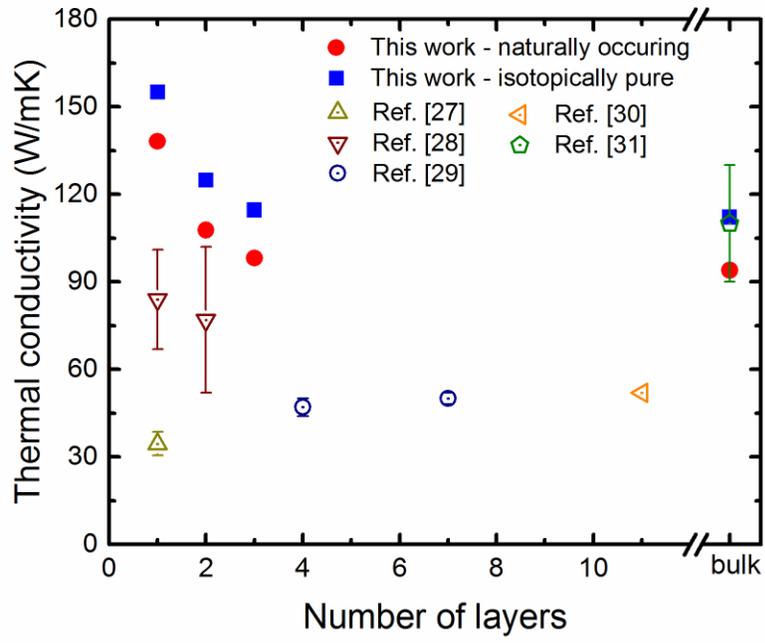

**Figure 7.**

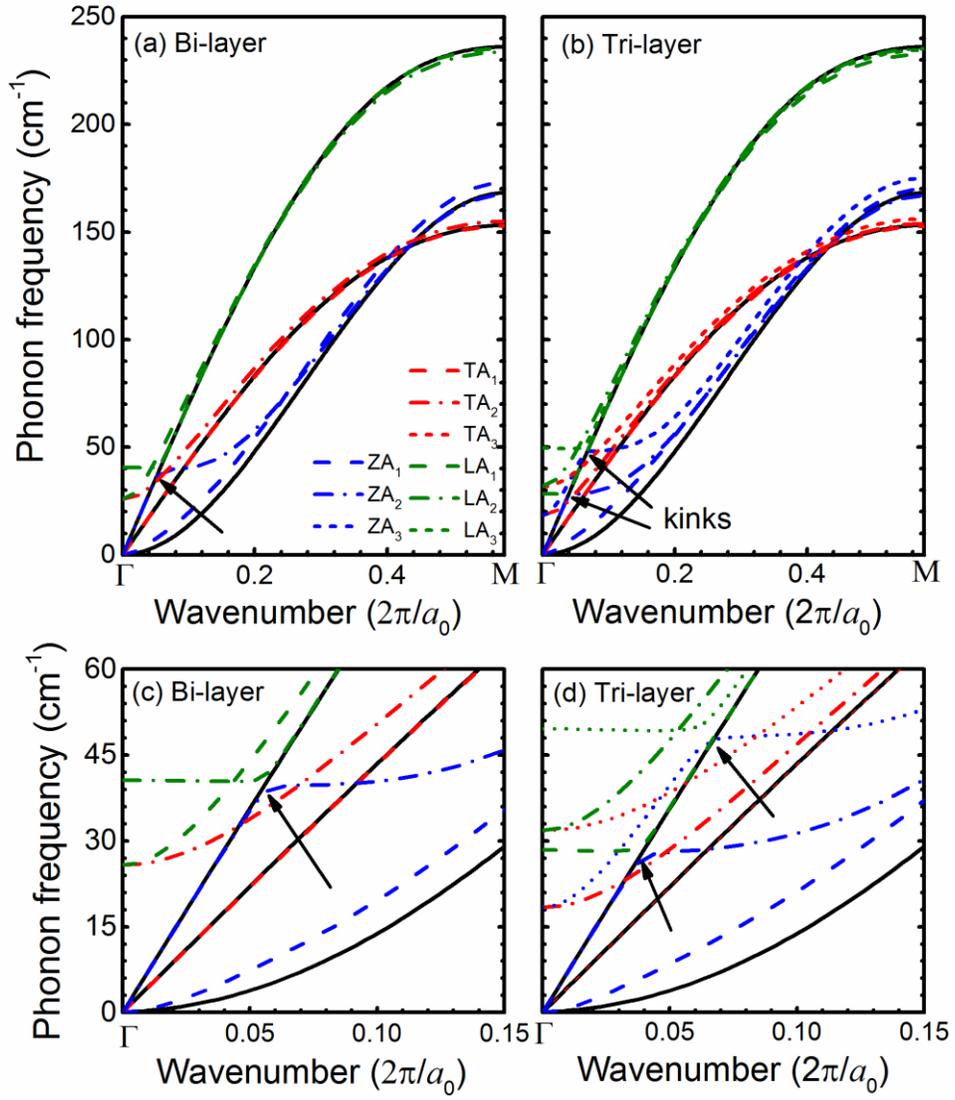

**Figure 8.**

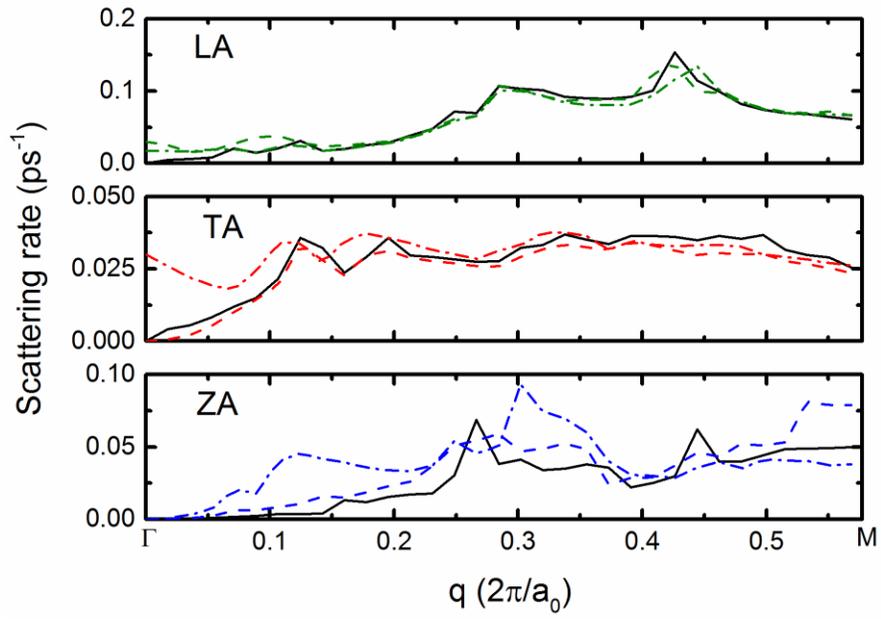